\documentclass[a4paper, usenatbib, useAMS, usegraphicx]{mn2e}

\usepackage{amssymb, color}

\usepackage[backref,breaklinks,colorlinks,citecolor=blue]{hyperref}
\usepackage[all]{hypcap}

\usepackage{color}
\newcommand{\red}[1]{{#1}}

\begin{document}

\title[Driving Galactic Turbulence]{Is Turbulence in the Interstellar Medium Driven by Feedback or Gravity? An Observational Test}

\author[Krumholz \& Burkhart]{
Mark R. Krumholz$^{1,2,3}$\thanks{mark.krumholz@anu.edu.au} and
Blakesley Burkhart$^4$\thanks{b.burkhart@cfa.harvard.edu}
\\ \\
$^1$Research School of Astronomy \& Astrophysics, Australian National University, Canberra, ACT 2611 Australia\\
$^2$Department of Astronomy \& Astrophysics, University of California, Santa Cruz, USA\\
$^3$Blaauw Visiting Professor, Kapteyn Astronomical Institute, University of Groningen, The Netherlands\\
$^4$Harvard-Smithsonian Center for Astrophysics, 60 Garden St., Cambridge, MA 0213
}

\maketitle

\begin{abstract}
Galaxies' interstellar media (ISM) are observed to be supersonically-turbulent, but the ultimate power source that drives turbulent motion remains uncertain. The two dominant models are that the turbulence is driven by star formation feedback and/or that it is produced by gravitational instability in the gas. Here we show that, while both models predict that the galaxies' ISM velocity dispersions will be positively correlated with their star formation rates, the forms of the correlation predicted by these two models are subtly but measurably different.  A feedback-driven origin for the turbulence predicts a velocity dispersion that rises more sharply with star formation rate, and that does not depend on the gas fraction (i.e. $\dot{M}_* \propto \sigma^2$), while a gravity-driven model yields  a shallower rise and a strong dependence on gas fraction(i.e. $\dot{M}_* \propto f_g^2 \sigma$). We compare the models to a collection of data on local and high-redshift galaxies culled from the literature, and show that the correlation expected for gravity-driven turbulence is a better match to the observations than a feedback-driven model. This suggests that gravity is the ultimate source of ISM turbulence, at least in the rapidly-star-forming, high velocity dispersion galaxies for which our test is most effective. We conclude by discussing the limitations of the present data set, and the prospects for future measurements to enable a more definitive test of the two models.
\end{abstract}


\section{Introduction}
\label{sec:intro}

The gas in galaxies is rarely if ever found in a quiescent disc supported purely by thermal pressure. Observations invariably reveal gas velocity dispersions indicative of supersonic motion. This is true for measurements in the local Universe using the 21 cm line of H~\textsc{i} \citep[e.g.,][]{van-zee99a, petric07a, tamburro09a, bls10,  ianjamasimanana12a, ianjamasimanana15a, stilp13a, cbls15}, the rotational transitions of CO \citep[e.g.,][]{pety13a, meidt13a, caldu-primo13a, caldu-primo15a}, or the recombination lines of ionized gas \citep[e.g.,][]{green14a, arribas14a, moiseev15a}. It is also true for galaxies at higher redshift measured in either CO or ionized gas \citep[e.g.,][]{cresci09a, lehnert09a, lehnert13a, green10a, green14a, le-tiran11a, swinbank12a, genzel14a}.

The observed large velocity dispersions provide strong evidence that the ISM is turbulent.
The turbulent nature of clouds in the Milky Way has been demonstrated by  a variety of observations including
the fractal and hierarchical structures of diffuse and molecular clouds \citep{elmegreen99, blgr13}, 
and the lognormal probability distribution functions (PDFs) of column density in the H$\alpha$ \citep{berkhuijsen08} and 21 cm lines \citep{bls10}, and in dust maps of
molecular clouds as measured from surveys such as Herschel and 2MASS \citep{Schneider15, burkhart15}.  In addition, a number of new 
techniques for diagnosing turbulence in the ISM have been developed, in particular to measure the 
velocity power spectrum \citep{cbls15} and the sonic and Alfv\'enic Mach numbers \citep{Kowal07, Burkhart09, Burkhart12, Esquivel11}. These methods have consistently shown that both diffuse and molecular gas in the Milky Way and in nearby local dwarf galaxies
is supersonic.

Despite the ubiquity of turbulence and superthermal velocity dispersions, the physical origin of these motions is far from clear \red{(e.g., see the recent review by \citealt{glazebrook13a}). Since the turbulent motions are present on a wide range of spatial scales, this question can be asked on a similarly wide range of scales, from individual clouds to the entire ISM. In this paper we focus on the largest, ISM-wide kiloparsec scales, though we note that there is also an active literature discussing the origins of turbulent motions \textit{within} individual molecular clouds on parsec scales \citep[e.g.,][]{krumholz06d, Ballesteros-Paredes11a, goldbaum11a, zamora-aviles14a, padoan15}.}

\red{On galactic scales, one obvious candidate for driving the motions} is stellar feedback. The giant H~\textsc{i} shells that have long been known in our own Galaxy \citep[e.g.,][]{heiles79a} provide clear evidence that young stars can drive significant motions in the neutral interstellar medium (ISM). Moreover, surveys of large numbers of galaxies indicate that the velocity dispersion in a galaxy's ISM is well-correlated with its star formation rate \citep[e.g.,][]{lehnert09a, lehnert13a, green10a, green14a, le-tiran11a, moiseev15a}, a correlation that several authors have interpreted as evidence that star formation feedback is the primary driver of turbulent motions.

However, theoretical models cast doubt on whether stellar feedback alone can explain the large velocity dispersions seen in more actively star-forming systems. Simulations using only supernova feedback show that they are capable of maintaining velocity dispersions of no more than $\sim 10$ km s$^{-1}$ \citep[e.g.,][]{dib06a, joung09a, shetty12a, kim13c, kim14b}. Reproducing the velocity dispersions of $\gg 10$ km s$^{-1}$ seen in some galaxies appears to require significantly more momentum input per unit mass of stars formed than the $\sim 3000$ km s$^{-1}$ SNe provide. Subgrid feedback models that assume higher rates of momentum injection (e.g., due to radiation pressure) can achieve high velocity dispersions \citep{hopkins11a, hopkins14a}, but it is unclear if such high momentum input rates are realistic \citep[e.g.,][]{krumholz12c, krumholz13a, davis14b, rosdahl15a, tsang15a}. Conversely, in very low surface density dwarf galaxies and the outskirts of spirals, where the star formation rate becomes very low, it seems difficult on energetic grounds to explain the presence of supersonic turbulence using supernova feedback alone \citep{stilp13a, padoan15}. 

The primary alternative possibility is that galaxies' non-thermal velocity dispersions are produced by some sort of instability. A number of authors have considered both thermal instability and magnetorotational instability as a potential driver of turbulence. However, numerical simulations of these mechanisms indicate that, while they do operate, both provide velocity dispersions of only a few km s$^{-1}$ \citep{kim03a, piontek04a, piontek05a, piontek07a, yang12a}. They cannot explain the large velocity dispersions of many tens of km s$^{-1}$ typically seen in rapidly star-forming, gas rich systems. Similarly, accretion of gas from the intergalactic medium appears insufficient on energetic grounds to power the observed velocity dispersions of galaxies over long timescales (e.g.~\citealt{elmegreen10a, hopkins13g}; however see \citealt{klessen10a} for a contrary view).

On the other hand, simulations indicate that gravitational instability can drive velocity dispersions as large as those observed in rapidly star-forming systems, and substantially larger than are produced by feedback alone \citep[e.g.,][]{bournaud07a, bournaud09b, agertz09a, ceverino10a, bournaud14a, goldbaum15a}. Analytic models, which appear quite consistent with the simulations, suggest that the instability creates enough turbulence to render the disc marginally stable, $Q\approx 1$, and that the energy to sustain this velocity dispersion is provided by accretion of mass through (as opposed to onto) the disc \citep{bertin99a, krumholz10c, forbes12a, forbes14a}.

While these results are suggestive, the gravitational instability explanation for the origin of the galaxies' velocity dispersions has yet to be tested against observations, and in particular against the observed correlations between galaxies' velocity dispersions, star formation rates, and gas fractions. Nor has the feedback-driven model been tested, except very roughly by noting the qualitative tend that velocity dispersion and star formation rate increase together. Our goal in this paper is to remedy this omission. In \autoref{sec:model}, we develop use a simple model for gravitational instability-dominated galaxies to derive a relationship between star formation, gas fraction, and velocity dispersion for such systems. We also develop a similar model for feedback-driven turbulence. We then compare both models to observations in \autoref{sec:observations}. We discuss our findings and their implications in \autoref{sec:discussion}.

\section{Model}
\label{sec:model}

\subsection{Gravity-Driven Turbulence}

To derive the expected relationship between gas content and star formation in a galactic disc where the turbulence is driven by gravity, we consider a system with a flat rotation curve with circular velocity $v_c$ and gas surface density and velocity dispersion versus radius $\Sigma$ and $\sigma$, respectively. The stellar surface density, considering only stars within $\sim 1$ gas scale height of the midplane, is $\Sigma_* = [(1-f_g)/f_g]\Sigma$, where $f_g$ is the gas fraction. 

For such a setup, \citet{krumholz10c} show that there exists a steady state configuration where turbulence is driven by gravity, ultimately powered by accretion through the disk. The steady state configuration is described by a family of similarity solutions where the \red{gas} surface density and velocity disperison versus radius are
\begin{eqnarray}
\label{eq:surfden}
\Sigma & = & \frac{v_c}{\pi G Q r} \left(\frac{f_g^2 G \dot{M}}{\eta}\right)^{1/3} \\
\sigma & = & \frac{1}{\sqrt{2}} \left(\frac{G\dot{M}}{\eta f_g}\right)^{1/3}.
\end{eqnarray}
Here $r$ is the galactocentric radius, $\dot{M}$ is the mass accretion rate through the disk, which is a free parameter, $\eta$ is a dimensionless number of order unity that measures the turbulent energy dissipation rate per scale height-crossing time, and
\begin{equation}
\label{eq:q}
Q \approx \sqrt{2} \frac{v_c\sigma f_g}{\pi G r \Sigma}
\end{equation}
is the gravitational stability parameter, which is constant in time and space. Note that $Q$ here is the total \citet{toomre64a} $Q$ parameter for both gas and stars, which we have approximated using the \citet{wang94a} approximation that $Q^{-1} \approx Q_g^{-1} + Q_*^{-1}$, where $Q_g$ and $Q_*$ are the Toomre parameters of the gas and stars alone (i.e., dropping the factor $f_g$ in the numerator of \autoref{eq:q}, and using $\Sigma$ or $\Sigma_*$ in the denominator, respectively). This distinction will become important below.

We now add star formation to this model. Observations over a wide range of scales show that the star formation rate per unit mass in molecular gas is well approximated by $\epsilon_{\rm ff}/t_{\rm ff}$, where $t_{\rm ff}$ is the free-fall time and $\epsilon_{\rm ff} \approx 0.01$ \citep{krumholz07e, krumholz12a, federrath13c, krumholz14c, salim15a}. Since the observations to which we are interested in comparing mostly consists of galaxies with high surface densities and molecule-dominated interstellar media, we need not consider the giant molecular cloud regime that prevails in Milky Way-like galaxies, wherein star-forming molecular clouds are far denser than the mean of the ISM. Instead, we consider only what \citet{krumholz12a} describe as the ``Toomre regime", where the entire ISM is a single star-forming structure, and the density relevant for star formation is simply the mid-plane density. In this regime, the free-fall time as a function of radius is
\begin{equation}
t_{\mathrm{ff}} = \sqrt{\frac{3}{2\phi_P}} \left(\frac{\pi Q r}{4 f_g v_c}\right)
\end{equation}
where $\phi_P \approx 3$ is a factor that accounts for the presence of stars in the disk. Using this in our similarity solution, the total star formation rate in a disk extending between radii $r_0$ and $r_1$ is
\begin{equation}
\label{eq:sfrvdisp_grav}
\dot{M}_* = \int_{r_0}^{r_1} 2\pi r \epsilon_{\mathrm{ff}} \frac{\Sigma}{t_{\mathrm{ff}}} \, dr
= \frac{16}{\pi} \sqrt{\frac{\phi_P}{3}} \left(\frac{\epsilon_{\rm ff} v_c^2}{G} \ln \frac{r_1}{r_0}\right) f_g^2 \sigma.
\end{equation}
Note that $Q$ and $\eta$ fall out of this relationship, so the only free parameters in the relationship between $\dot{M}_*$ and $\sigma$ are the mid-plane gas fraction $f_g$ and a Coulomb logarithm-like term $\ln (r_1/r_0)$, which measures the radial extent of the star-forming disk.

Although the values of $r_0$ and $r_1$ obviously enter only logarithmically, numerical evaluate requires that we have least rough estimates for them. At small radii the analytic solution we are using breaks down because $v_c$ cannot remain constant all the way to $r=0$. Instead, the rotation curve must turn over. A rotation curve with constant $v_c$ provides an infinite amount of shear as $r\rightarrow 0$, making it possible to maintain an infinitely large gas surface density at fixed $Q$, which is why the contribution to the total star formation rate from small radii diverges as $r_0\rightarrow 0$. Since a finite amount of shear will support only a finite surface, in reality this divergence will not occur, which suggests that the natural choice for $r_0$ is the radius at which the rotation curve turns down from being flat. In the Milky Way this occurs at $r_0 \approx 100$ pc \citep{launhardt02a, krumholz15d}, and we adopt this as a fiducial value.\footnote{The gas distribution in the Milky Way is modified at small radii by the presence of the galactic bar, but the majority of the galaxies with which we will compare theoretical models below are very gas-rich and thus likely lack bars, so we do not consider bars further here.}

The outer radius $r_1$ will be determined by the edge of the star-forming disk, where the ISM transitions from molecular- to atomic-dominated, and the gas depletion time increases by $\sim 1-2$ orders of magnitude \citep{bigiel10a, krumholz14c}. This radius varies from galaxy to galaxy, but it is always within an order of magnitude of 10 kpc. Since high precision is not required here due to the logarithmic dependence, we simply adopt $r_1 = 10$ kpc as a fiducial value for simplicity.\footnote{Note that the choice of $r_1$ is equivalent, for a particular choice of $f_g$, $Q$, $\eta$, and $\sigma$, to choosing the total molecular gas mass (via integration of \autoref{eq:surfden}). Specifically, for $Q = 1$ and $\eta=3/2$ (as recommended by \citealt{krumholz10c}), we have $M_{\rm gas} = 1.2\times 10^{10} f_g \sigma_{10} v_{c,200} r_{1,10}$ $M_\odot$, where $\sigma_{10} = \sigma/(10\,\mathrm{km}\,\mathrm{s}^{-1})$, $v_{c,200} = v_c / (200\,\mathrm{km}\,\mathrm{s}^{-1})$, and $r_{1,10} = r_1/(10\,\mathrm{kpc})$.} With this choice, we have fully determined the relationship between $\dot{M}_*$ and $\sigma$ expected if turbulence is driven by gravity.

\subsection{Feedback-Driven Turbulence}

Numerous authors have proposed models of feedback-driven turbulence \citep[e.g.,][]{thompson05a, ostriker11a, shetty12a, faucher-giguere13a}, and they make a range of predictions. Broadly speaking, the models can be classified into two approaches. One is to assume that the star formation rate is dictated solely by the requirement that momentum injection from star formation balance gravity, while the dimensionless star formation rate per free-fall time $\epsilon_{\rm ff}$ remains approximately constant, as observations seem to suggest \citep[e.g.,][]{ostriker11a, shetty12a}. In these models, the velocity dispersion varies weakly or not at all with the properties of the galaxy. Consequently, these models do not require the enforcement of $Q\approx 1$. While these results are consistent with simulations indicating that supernova-driven turbulence cannot produce large velocity dispersions, they cannot explain the large observed velocity dispersions that are the focus of this paper, and so we do not discuss them further.

The alternative approach \citep[e.g.,][]{thompson05a, faucher-giguere13a} is to assume that the star formation rate for fixed gas properties, as parameterized by $\epsilon_{\rm ff}$, is an extremely steep function of $Q_g$. \red{We pause here to note a subtle but important point. In these models, the key physics that regulates galaxy disks is the rate at which bound structures -- giant molecular clouds (GMCs) -- form out of the diffuse ISM. Within these structures, $\epsilon_{\rm ff}$ is assumed to be very high\footnote{It is unclear if the \red{high values of} $\epsilon_{\rm ff}$ required in these models can be reconciled with observations \citep{krumholz14c}, but we put this issue aside and focus on the key prediction that can be tested from velocity dispersions}, and thus the rate-limiting step in star formation is the rate of GMC formation. Since the stellar potential is relatively smooth on the scales of GMCs, this process is driven by the self-gravity of the gas alone, not by the gravity of the stars. Thus the rate of star formation in these models depends not on $Q$ but on $Q_g$, and it does so in an almost step function-like way. Star formation, and thus the momentum injection from stars, goes to zero rapidly at $Q_g > 1$, even if $Q \approx 1$. Since in these models the star formation rate is in turn dictated by the requirement that this momentum injection maintain the turbulence in the ISM,} these models enforce $Q_g \approx 1$, \red{not} $Q \approx 1$. Thus they predict (c.f., equation \red{6} of \citealt{faucher-giguere13a})
\begin{equation}
\label{eq:sigma_Qg}
\sigma = \frac{\pi G r \Sigma Q_g}{\sqrt{2} v_c} \approx \frac{\pi G r \Sigma}{\sqrt{2} v_c},
\end{equation}
whereas gravitationally-driven turbulence would introduce an extra factor $f_g$ in the denominator.

\red{Once $Q_g$ reaches unity and star formation turns on, its} rate in these models is determined by the requirement to maintain hydrostatic balance, which implies a star formation rate that varies inversely with the momentum supplied per unit mass of stars formed, and directly as the square of the gas surface density. To be definite we adopt the relation derived by \citet[their equation 18]{faucher-giguere13a}, but, as noted in their paper, this result is essentially the same in all feedback-driven turbulence models where $Q_g$ is kept fixed rather than $\epsilon_{\rm ff}$. This relationship is
\begin{equation}
\label{eq:sigmasfr}
\dot{\Sigma}_* = \frac{2\sqrt{2}\pi G Q_g \phi}{\mathcal{F}} \left(\frac{P_*}{m_*}\right)^{-1} \Sigma^2,
\end{equation}
where $\phi \approx 1$ and $\mathcal{F} \approx 2$ are constants of order unity that parameterize various uncertainties; the numerical values given here are the ones recommended by \citet{faucher-giguere13a}. The quantity $P_*/m_*$ is the momentum injected per unit mass of stars formed, for which we also adopt \citet{faucher-giguere13a}'s recommended value of 3000 km s$^{-1}$. If we now adopt a value of $\sigma$ that is independent of radius, in rough agreement with the observed weak variation of $\sigma$ within galaxies, and use \autoref{eq:sigma_Qg} to eliminate $\Sigma$, we can integrate \autoref{eq:sigmasfr} with radius to obtain the relationship between star formation rate and gas velocity dispersion,
\begin{equation}
\label{eq:sfrvdisp_fb}
\dot{M}_* = \int_{r_0}^{r_1} 2\pi r \dot{\Sigma}_*\, dr = \frac{8\sqrt{2} \phi v_c^2}{\pi G Q_g \mathcal{F}}\left(\ln\frac{r_1}{r_0}\right) \left(\frac{P_*}{m_*}\right)^{-1} \sigma^2.
\end{equation}

Comparing \autoref{eq:sfrvdisp_grav} to \autoref{eq:sfrvdisp_fb}, we see that gravity-driven turbulence models predict $\dot{M}_* \propto f_g^2 \sigma$, while feedback-driven models give $\dot{M}_* \propto \sigma^2$, with no dependence on $f_g$. In \autoref{sec:observations} we will use these differences to test the models against observations, but first we pause to understand the physical origins of the different scalings, which are two-fold.

First, as noted above, gravity-driven models enforce that $Q \approx 1$, because the total $Q$ parameter is what controls the strength of the gravitational instability that drives turbulence. In contrast, feedback-driven models require $Q_g \approx 1$, because  the amount of gas that is unstable to collapse, and thus the amount of momentum injected by feedback, depends so sharply on $Q_g$ that $Q_g$ stays about constant regardless of the value of $Q$.\footnote{In the terminology of \citet{faucher-giguere13a}, this is formally their $\alpha\rightarrow \infty$ limit.} Thus, for example, consider what happens if the stellar content of a galaxy decreases while the gas content is held constant, causing $f_g$ to rise. In this case a gravity-driven model predicts that the the gas velocity dispersion will fall, because less turbulence is needed to maintain stability. In contrast, in feedback-dominated models the velocity dispersion remains essentially unchanged.

The second origin of the difference in scaling is the relationship between gas and star formation surface densities. When feedback is assumed to drive the turbulence that balances the weight of the ISM, the rate of momentum injection, and thus star formation, must vary as the square of the gas surface density. In contrast, when gravitational instability is assumed to be the source of the turbulence, there is no need for the star formation rate to rise so sharply with gas content, and instead the scaling is only $\dot{\Sigma}_* \propto \Sigma$ for fixed galactic angular velocity. Since gas surface density $\Sigma$ and velocity dispersion $\sigma$ are proportional to one another at fixed $Q$, this difference implies a star formation rate that rises as $\sigma^2$ for feedback-driven turbulence, versus one that rises as $\sigma$ for gravity-driven turbulence.

\section{Comparison to Observations}
\label{sec:observations}

\subsection{Data}
\label{ssec:data}

We now compare the predictions of the theoretical models to a set of observations culled from the literature. The data include measurements of the star formation rate and velocity dispersion in either ionized gas or H~\textsc{i}. Where possible we have also obtained gas fractions, in order to test the dependence of the $\sigma-\dot{M}_*$ relationship on gas fraction that is predicted to exist for gravity-driven models but not feedback-driven ones. We use the following data sources for our velocity dispersions: H$\alpha$ measurements of local spirals \citep{epinat08a} and dwarfs \citep{moiseev15a}; H$\alpha$ measurements at $z\sim 0.1$ \citep{green10a} and at a range of $z\sim 1 - 3$ \citep{epinat09a, cresci09a, law09a, lemoine-busserolle10a, lehnert13a}; and H~\textsc{i} measurements of local galaxies \citep{ianjamasimanana12a}. For the H$\alpha$ data, the star formation rates come from the H$\alpha$ line and are reported in the papers as the kinematics. Star formation rates to go with the H~\textsc{i} kinematics are from \citet{leroy08a} or, if that is unavailable, the literature compilation of \citet{walter08a}. We take gas fractions for the samples of \citet{law09a} and \citet{lemoine-busserolle10a} from the values reported by those authors, while gas fractions for all other $z>0$ galaxies, where they are available, come from \citet{tacconi13a}. For the local galaxies, we compute gas fractions from \citet{leroy08a}'s reported H~\textsc{i}, H$_2$, and stellar masses.

Before making this comparison, we must account for an effect that is present in the observations but not in the models: with only one exception, the velocity dispersion measurements to which we have access are for ionized gas, meaning that we are measuring the gas inside H~\textsc{ii} regions. This gas will have a minimum velocity dispersion of $\approx 10$ km s$^{-1}$ simply as a result of thermal broadening. Moreover, H~\textsc{ii} regions expand at velocities up to $\sim 10$ km s$^{-1}$ for small regions, and at up to $\sim 30-40$ km s$^{-1}$ for the largest ones \citep[e.g., 30 Doradus --][]{chu94a}. \red{The H$\alpha$ spectrum that we observe from any individual aperture, which in extragalactic work almost always contains several H~\textsc{ii} regions, will therefore be a sum of roughly Gaussian profiles, each of which has an intrinsic width of $\sim 10$ km s$^{-1}$ or more. The quantity of interest to us is the velocity dispersion in the neutral ISM, which should be roughly the dispersion the centroids of these Gaussians. However, in cases where the dispersion of the centroids is smaller than $\sim 10$ km s$^{-1}$, the intrinsic width coming from thermal and non-thermal motions within H~\textsc{ii} regions will set a lower limit on the value we actually measure for the dispersion of the spectrum.}
Some authors attempt to correct for the thermal broadening \citep[e.g.,][]{moiseev15a}, but most do not, and none attempt to correct for the expansion broadening.  The extent to which either effect influences the authors' reported results depends on the exact procedure that the authors use to determine the velocity dispersion, but at measured velocity dispersions of order $\sim 10$ km s$^{-1}$, the effect is non-negligible, as is apparent from the fact that the velocity dispersions obtained by \citet{ianjamasimanana12a} using H~\textsc{i} are systematically a factor of $\sim 2$ smaller than the ionized gas measurements of \citet{epinat08a} and \citet{moiseev15a} at similar redshifts and star formation rates.

Rather than attempting to correct the heterogenous data set that we have assembled to some sort of uniformity, we have elected simply to use the values reported by the authors, and to add a correction to the theoretical predictions to account to H~\textsc{ii} region expansion. Thus we add a velocity dispersion of 15 km s$^{-1}$ in quadrature to the value predicted by the theoretical models in all the figures below. Since, as we shall see, the majority of the power of the observations in distinguishing between models comes from galaxies with $\sigma \gg 15$ km s$^{-1}$, the effects of this correction are small.

\subsection{The $\dot{M}_* - \sigma$ Relation}

\begin{figure}
\includegraphics[width=8.5cm]{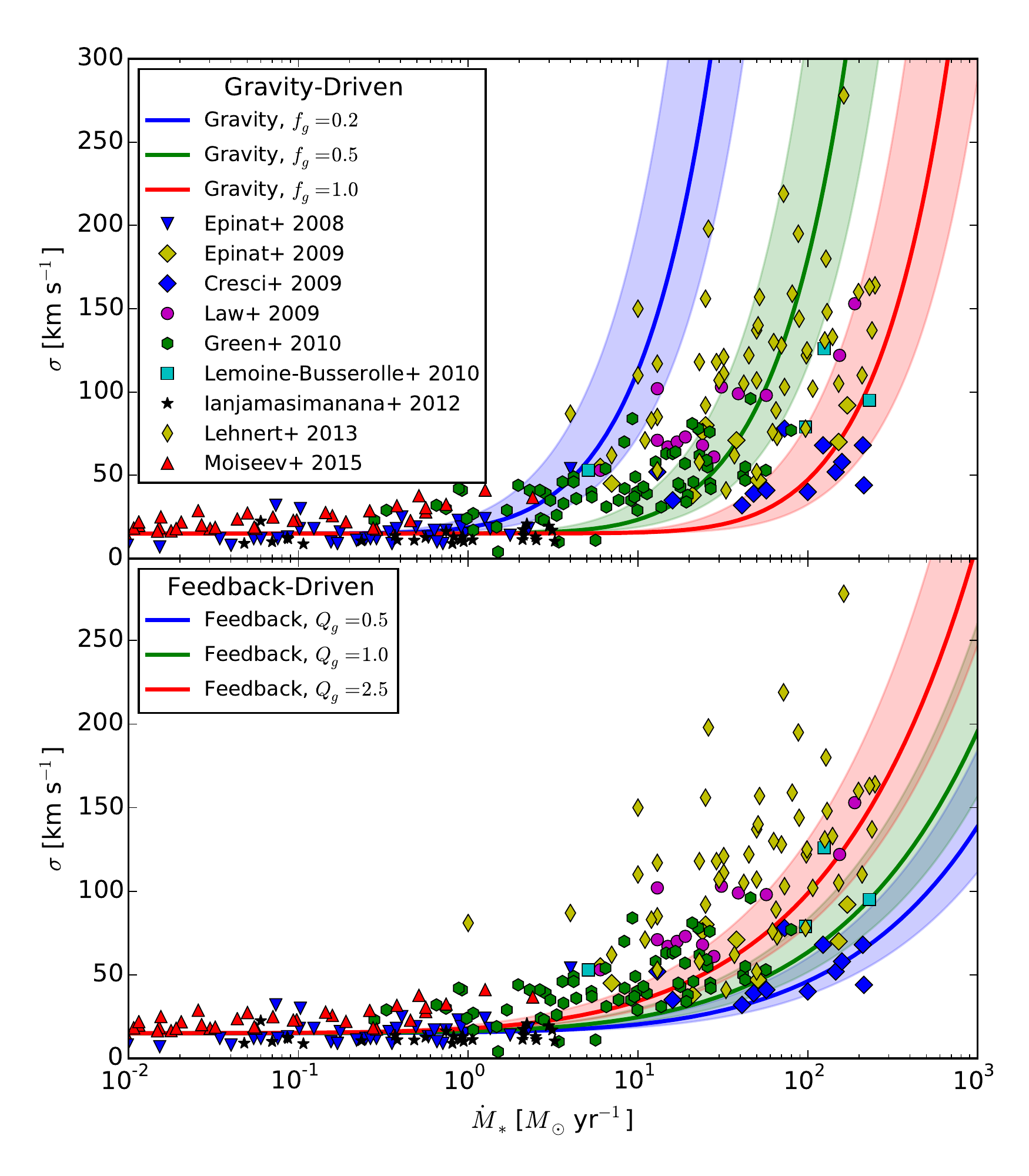}
\caption{
\label{fig:alldata}
The relationship between star formation rate, $\dot{M}_*$, and velocity dispersion, $\sigma$. In the top panel, lines show the predictions of the gravity-driven model (\autoref{eq:sfrvdisp_grav}) for $f_g = 0.2, 0.5$, and $1.0$, as indicated in the legend. Lines in the bottom panel show the prediction of the feedback-driven model (\autoref{eq:sfrvdisp_fb}) for $Q_g = 0.5, 1.0$, and $2.5$. The solid lines are for a circular velocity $v_c = 200$ km s$^{-1}$, and the shaded range shows values from $v_c = 150 - 250$ km s$^{-1}$, with larger values of $v_c$ corresponding to smaller values of $\sigma$. Note that the theoretical model predictions for $\sigma$ have been added in quadrature with 15 km s$^{-1}$ -- see \autoref{ssec:data}. Points show observations, from the sources indicated in the legend, and are the same in both panels.
}
\end{figure}

In \autoref{fig:alldata} we show the relationship between star formation rate and velocity dispersion as revealed by the data, and compare to the predictions of the gravity-dominated (\autoref{eq:sfrvdisp_grav}) and feedback-dominated (\autoref{eq:sfrvdisp_fb}) models. For the theoretical models, we show a range of $f_g$ (for the gravity models) and $Q_g$ (for the feedback models) that spans the plausible range for the observations. We also show results for circular velocities $v_c = 150 - 250$ km s$^{-1}$, which spans the plausible range for the high-redshift sample. The local dwarfs may go to smaller $v_c$, but they lie in a part of the plot where the models predict low velocity dispersions regardless of the value of $v_c$, so we focus on the range applicable to the higher star formation and velocity dispersion part of the sample.

The data clearly reveal that $\sigma$ increases with $\dot{M}_*$, as noted before by numerous authors, and as some have argued demonstrates that turbulence must be driven by stellar feedback. However, we see that the quantitative predictions of a stellar feedback-driven model provide rather a poor match to the observations. In particular, there are numerous galaxies with star formation rates of $\sim 10-100$ $M_\odot$ yr$^{-1}$ and velocity dispersion $\gtrsim 50$ km s$^{-1}$. Such a combination is extremely difficult to arrange in a feedback-dominated model, for a simple physical reason: if the gas is required to remain near $Q_g \approx 1$, then a high velocity dispersion implies a high gas surface density. However, in a feedback-driven model a high gas surface density necessitates a very high star formation rate, due to the steep $\dot{\Sigma}_*\propto \Sigma^2$ dependence implied by such models. The consequence is that velocity dispersion does not rise steeply enough with star formation rate to match the data.

The gravity-driven model, on the other hand, shows far better agreement with the observations. The general shape of $\sigma$ versus $\dot{M}_*$ predicted by gravity-driven turbulence matches the trend in the data. In particular, we see that the gravity-driven model has no trouble reproducing low high values of $\sigma$ at relatively modest star formation rates, interpreting these cases as galaxies where the gas fraction is relatively low. The low gas fraction suppresses the star formation rate, but the velocity dispersion remains high due to the need to keep the entire galaxy near $Q\approx 1$.

\subsection{Dependence on the Gas Fraction}
\label{ssec:gasfrac}

\begin{figure}
\includegraphics[width=8.5cm]{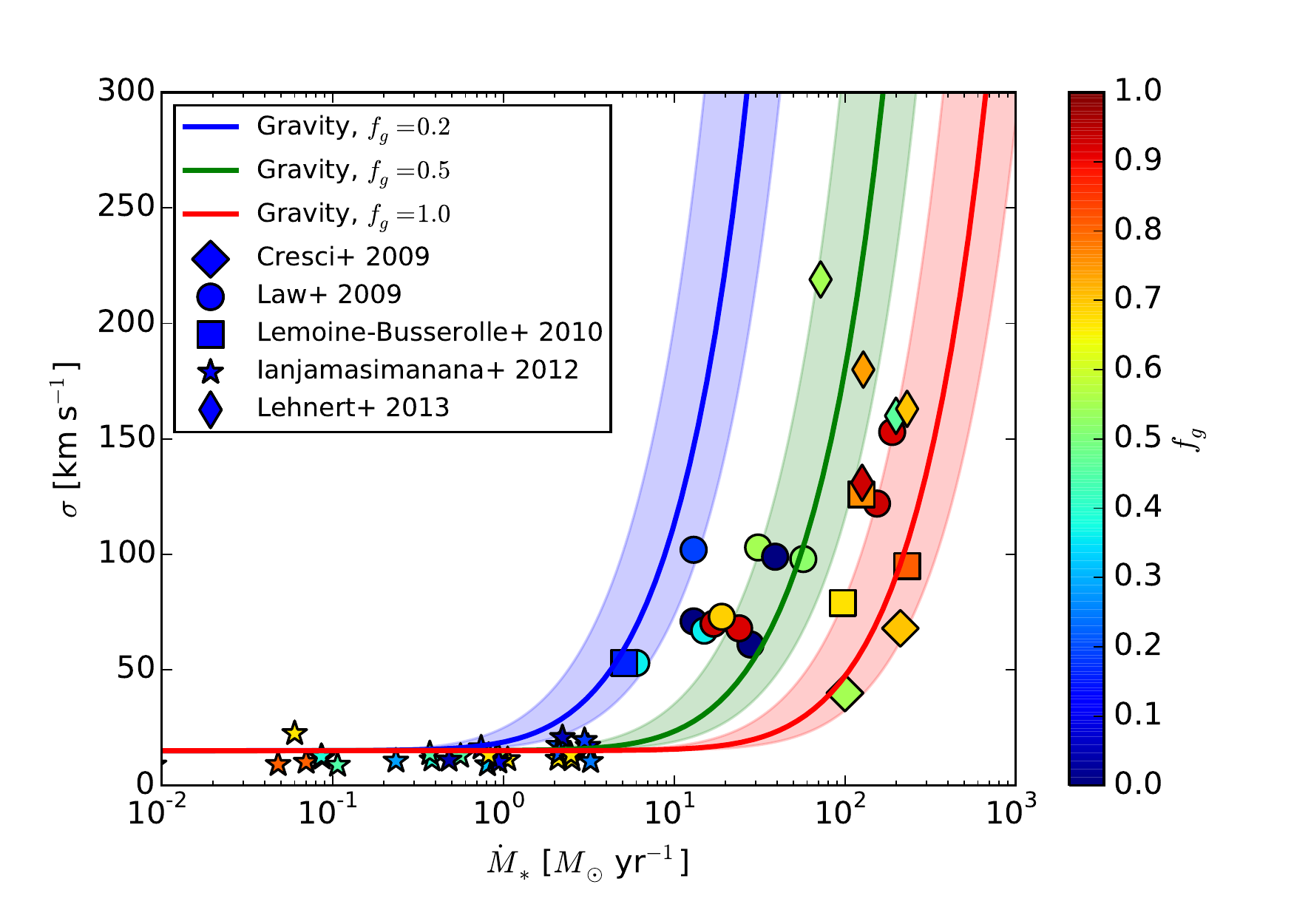}
\caption{
\label{fig:fgdata}
Same as the top panel of \autoref{fig:alldata}, but now showing only observations for which a gas fraction is available, and with observed points color-coded by gas fraction, from $f_g \approx 0$ (blue) to $f_g \approx 1$ (red). Data sources are as indicated in the legend, and theoretical models are the same as in the top panel of \autoref{fig:alldata}.
}
\end{figure}

\begin{figure}
\includegraphics[width=8.5cm]{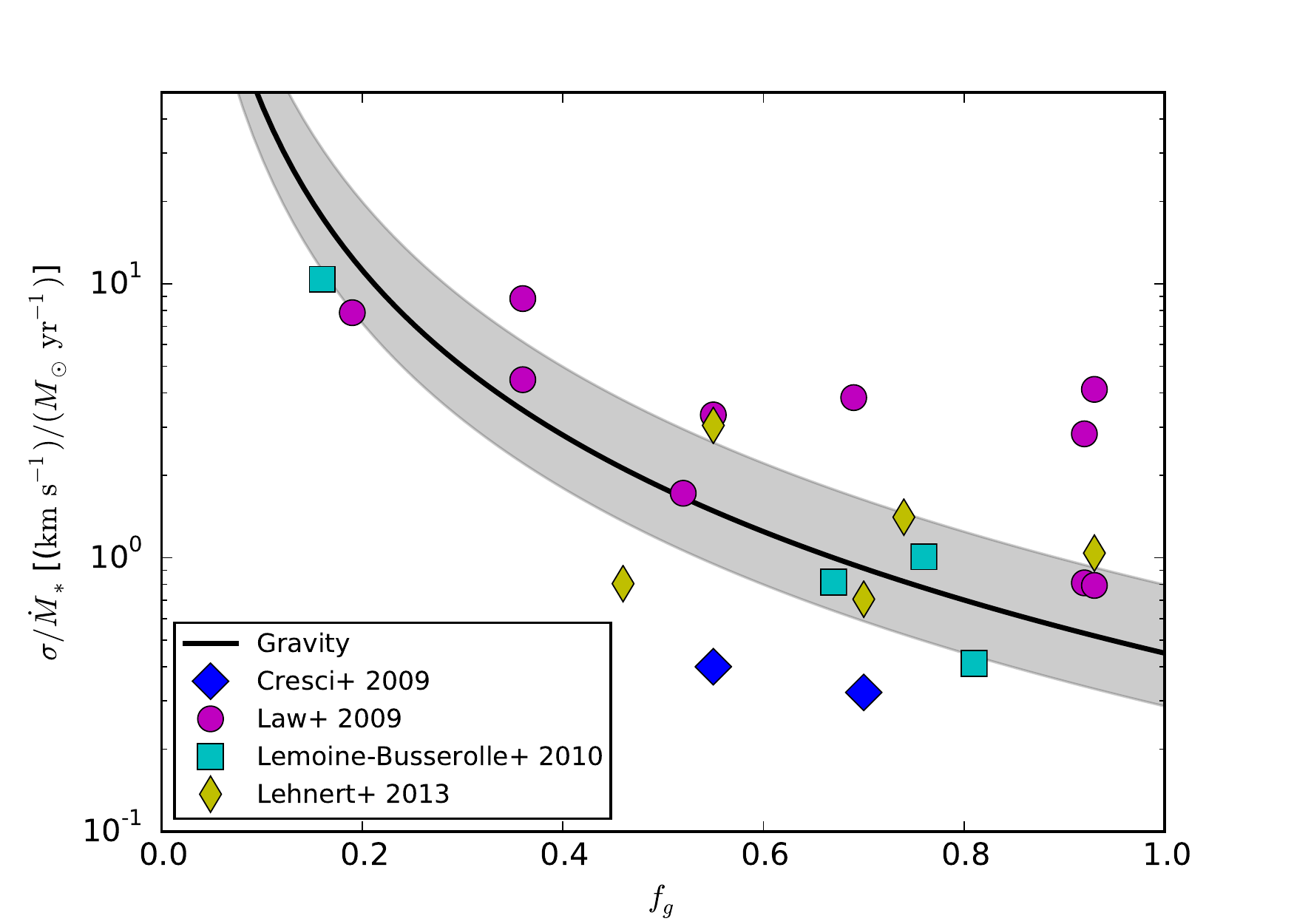}
\caption{
\label{fig:fgcomp}
Ratio of velocity dispersion to star formation rate, $\sigma/\dot{M}_*$, versus gas fraction, $f_g$. Data points are from the sources indicated in the legend. The black line shows the prediction of the gravity-driven model (\autoref{eq:sfrvdisp_grav}) evaluated with $v_c = 200$ km s$^{-1}$, while the band indicates the range from $v_c = 150 - 250$ km s$^{-1}$. Unlike in \autoref{fig:alldata} and \autoref{fig:fgdata}, in this plot we do not add the 15 km s$^{-1}$ correction to the theoretically-predicted velocity dispersion, because the model predicts a unique value of $\sigma/\dot{M}_*$, but not $\sqrt{\sigma^2 + (15\,\mathrm{km}\,\mathrm{s}^{-1})^2}/\dot{M}_*$.
}
\end{figure}

We next investigate whether the relationship between $\dot{M}_*$ and $\sigma$ depends on gas fraction, as predicted by gravity-driven models but not by feedback-driven ones. This requires that we have access to the gas fraction, which is available only for a relatively small subset of the data. Moreover, the gas fractions we do have available are global measurements, whereas the quantity of interest to us is the gas fraction at the galactic midplane, which could be systematically different.

With these caveats in mind, in \autoref{fig:fgdata} we show $\sigma$ versus $\dot{M}_*$ with the data points color-coded by the measured gas fraction. While there is a great deal of scatter, there is a clear systematic trend between gas fraction and location in the $\dot{M}_* - \sigma$ plane: galaxies with \red{higher} gas fractions tend to fall to the lower right of the diagram, at high star formation rate and low velocity dispersion, while those with low gas fraction tend toward low star formation rate and high velocity dispersion. This is precisely the trend predicted by the gravity-driven models.

To make this comparison, more clearly in \autoref{fig:fgcomp} we directly compare the value of $\sigma/\dot{M}_*$ to $f_g$.\footnote{We omit the local data set from \citet{ianjamasimanana12a} here, because all of these measurements lie in the region where their velocity dispersions are dominated by the thermal correction, and are essentially the same. The remaining data that we retain have velocity dispersions for which this correction is negligible.} The models predict a systematic decrease in $\sigma/\dot{M}_*$ with $f_g$, and, while there is a great deal of scatter, such a trend does appear in the data. In contrast, recall that a feedback-driven model for the origin of the turbulence would predict no trend between $\sigma/\dot{M}_*$ and $f_g$. While the data are very noisy, they are clearly better described by a value of $\sigma/\dot{M}_*$ that varies with $f_g$ than one that is constant with $f_g$.

\section{Discussion and Summary}
\label{sec:discussion}

\subsection{Gravity-Driven Turbulence}

The most basic result of this work is that the relationship between star formation rate, velocity dispersion, and gas fraction represents a useful test of the physical origins of interstellar turbulence, and that the existing data appear to favor a model in which the turbulence \red{in galaxies with star formation rates above a few $M_\odot$ yr$^{-1}$} originates from gravitational instability rather than stellar feedback. \red{At lower star formation rates, the velocity dispersions measured from ionized gas are dominated by the internal motions of H~\textsc{ii} regions, and it is not possible to distinguish the models given the existing data set. At higher star formation rates,} there are two main lines of evidence for \red{a gravitational origin for the motions}.

First, a feedback-dominated model predicts a rather steep dependence of the star formation rate on the gas surface density, and since surface density and velocity dispersion are linked via \citet{toomre64a}'s $Q$, on the velocity dispersion.\footnote{One might think that this prediction could be tested more naturally via a direct measurement of the correlation between star formation rate and gas surface density. However, this measurement is subject to numerous uncertainties regarding how one converts measured CO luminosity to mass \citep{bolatto13a} -- for example see \citet{faucher-giguere13a} versus \citet{genzel15a}. For this reason, the correlation between gas fraction and velocity dispersion may actually be a stronger test, since velocity dispersion does not suffer from uncertainties regarding conversion factors.} Consequently, velocity dispersions of $\sim 50$ km s$^{-1}$ can be driven only by star formation rates $\gtrsim 100$ $M_\odot$ yr$^{-1}$. In contrast, if the energy used to drive the turbulence comes from gravity rather than stellar feedback, there is no need for such a steep dependence of star formation rate on gas velocity dispersion. In this case it becomes possible to have high velocity dispersions of $\sim 50$ km s$^{-1}$ even at star formation rates of $\sim 10$ $M_\odot$ yr$^{-1}$, if the gas fraction is low enough -- the gravity of the stars still provides the energy needed to maintain high $\sigma$, while the paucity of gas keeps the star formation rate low. Observations indicate that there are significant numbers of galaxies in the range of parameter space that is forbidden to feedback-dominated models, but allowed for gravity-dominated ones.

The second line of evidence comes from the relationship between velocity dispersion and gas fraction. Because gravitational instability can be driven by stars as well as gas, galaxies with low gas fractions tend to have comparatively higher velocity dispersions (at fixed star formation rate) than galaxies with high gas fractions. The result is a negative correlation between $\sigma/\dot{M}_*$ and gas fraction. In contrast, feedback-driven models predict no such correlation. While the data are sparse and noisy, they appear to be more consistent with the negative correlation predicted by gravity-dominated models than with the lack of variation predicted if feedback is the mechanism driving turbulence.

A final, higher level conclusion is worth taking away from this work as well. A number of authors have argued that a correlation between star formation rate and velocity dispersion provides evidence that feedback drives turbulence. However, we have shown that such a correlation emerges generically in almost any plausible model of the origin of galaxy-scale turbulence. The correlation by itself is evidence of little except that galaxies with more gas in them tend to have both higher velocity dispersions and more star formation. A more quantitative approach is needed, including investigations of how the local star formation rate varies with local gas dispersion across individual galaxies \citep[e.g.,][]{tamburro09a}.

\subsection{Caveats and Future Tests}

While the data we have compiled are suggestive, they cannot be considered definitive. We mention two caveats here that seem particularly pressing, and that point out directions for future work. First, the high velocity dispersions at moderate star formation rates that are most powerful for discriminating between the two models come primarily from $z\gtrsim 1$ galaxies where resolution is limited and beam smearing is a possible concern\red{, i.e., where it might be difficult to disentangle turbulent motions within the ISM from the overall rotation of the galaxy}. This is not entirely true -- some of the points that lie outside the envelope of the gravity-driven models come from the local sample of \citet{epinat08a} and the $z\sim 0.1$ data set of \citet{green10a}, where beam smearing is much reduced. Nonetheless, our tentative conclusions could be placed on much more solid footing (or invalidated) by an expanded set of galaxies at low redshift with moderate star formation rates ($\sim 5-30$ $M_\odot$ yr$^{-1}$).  In addition, measurements of the velocity power spectrum in local galaxies (e.g., using the Velocity Coordinate Spectrum technique) would be useful for obtaining the driving scale associated with supernova driving versus gravitational driving of turbulence \citep{cbls15}.

A second obvious caveat is the paucity of gas fraction measurements. The strongest tests of the effects of gas fraction can be found in the regime of high velocity dispersion where the contribution from local expansion velocities of H~\textsc{ii} regions is small enough not to be problematic (\autoref{fig:fgcomp}). However, we have only $\sim 20$ measured gas fractions in this regime, and the measurement is an intrinsically noisy one for the reasons discussed in \autoref{ssec:gasfrac}. It is clearly a high priority to obtain more gas fraction measurements in high velocity dispersion galaxies. Again, an obvious target is galaxies at modest redshift, where the investment of telescope time required to obtain a large sample of molecular line measurements is less daunting than for the high-$z$ sample.

\section*{Acknowledgements}

MRK thanks F.~Fraternali for helpful discussions. MRK is supported by grant DP160100695 from the Australian Research Council and grant AST-1405962 from the US NSF. BB  is supported by the NASA Einstein Postdoctoral Fellowship.

\bibliographystyle{mn2e}
\bibliography{refs}

\end{document}